\documentclass[12pt]{iopart}
\usepackage{color}
\usepackage{iopams}
\usepackage{setstack}
\usepackage{bm}
\usepackage{graphicx}

\begin{document}

\title[Design principle of multi-cluster and desynchronized states in
oscillatory media]
{Design principle of multi-cluster and desynchronized states in
oscillatory media via nonlinear global feedback}

\author{Yasuaki Kobayashi$^1$ and Hiroshi Kori$^2$}
\address{$^1$Meme Media Laboratory, Hokkaido University, Sapporo
060-0813, Japan}
\address{$^2$Division of Advanced Sciences, Ochadai Academic
Production, Ochanomizu University, Tokyo 112-8610, Japan}
\eads{\mailto{$^1$kobayashi@nsc.es.hokudai.ac.jp}, \mailto{$^2$kori.hiroshi@ocha.ac.jp}}

\begin{abstract}
A theoretical framework is developed for a precise control of spatial
patterns in oscillatory media using nonlinear global feedback,
where a proper form of the feedback function corresponding to a
specific pattern is predicted through the analysis of a phase
diffusion equation with global coupling.
In particular, feedback functions that generate the following spatial
patterns are analytically given: i) 2-cluster states with an arbitrary
population ratio, ii) equally populated multi-cluster states, and iii)
a desynchronized state.
Our method is demonstrated numerically by using the Brusselator model
in the oscillatory regime.
Experimental realization is also discussed.
\end{abstract}



\section{Introduction}
Feedback control is a powerful method of regulating spatio-temporal
dynamics and has been studied in a wide variety of fields including
physics, chemistry, biology, and medical science \cite{mikhailov06}.
For example, formation of various clustering patterns has been realized in
the Belousov-Zhabotinsky reaction \cite{vanag00, vanag00_th_exp} and in
the catalytic CO oxidation reaction on Pt \cite{pollmann01,
bertram03-ex, kim01}.  
The catalytic CO oxidation systems have also been studied for the
suppression of chemical turbulence \cite{bertram03-ex, beta03}. 
Moreover, considerable attention has been paid to feedback devices
that suppress the pathological synchronization in the brain
of Parkinson's disease patients \cite{popovych05, hauptmann05,
popovych06, tukhlina07, rosenblum04, rosenblum04_2}. 

In many cases systems to be controlled are spatially extended, and
reaction-diffusion systems provide a good model for the study of pattern
controlling. 
Theoretical analyses based on reaction-diffusion systems have been done
for the Belousov-Zhabotinsky reaction  \cite{vanag00_th_exp, yang00, rotstein03} and CO
oxidation \cite{bertram03-the, beta04, parmananda01}. 
However, so far, only empirical control has been achieved for
such spatially-extended systems, including above-mentioned pioneering
experimental works \cite{vanag00, vanag00_th_exp, kim01}; there has been
no general theory that {\em quantitatively} relates feedback inputs to spatial
patterns. 

On the other hand, for discrete oscillator systems, such a quantitative
feedback control methodology has been 
established very recently by Kiss, Kori, Hudson, and Rusin \cite{kiss07,
kori08}.  
Their method is based on a phase model described by 
\begin{equation}
 \frac{\rmd \phi_i}{\rmd t} = \omega + \frac{K}{N}\sum_{j=1}^N \Gamma(\phi_i-\phi_j),
  \label{eq:pm}
\end{equation}
where $\phi_i$ ($0 \le \phi_i < 2\pi$) is the phase of the oscillator $i$
($i=1,\ldots,N$), $\omega$ is the natural frequency, $K$ is the coupling
strength, and $\Gamma(\phi)$
is called the coupling function. 
Their method utilizes the following facts: the coupling function determines the
entire collective behavior of the phase model, and any coupling
function can be designed by applying an appropriately constructed
feedback signal to a population of oscillators. 
Hence the population of oscillators can be
steered to a desired synchronization behavior by taking the following
two steps: (i) find a coupling function that results in a desired
synchronization behavior in (1), and (ii) construct an appropriate
feedback signal that yields the coupling function. 
A major advantage of their methodology is that the phase model can be
constructed from experimentally 
measurable quantities only; detailed information on the
intrinsic dynamics of the system is not necessary. 
Validity and robustness of their methodology have been confirmed
both experimentally by using electro-chemical oscillators \cite{kiss07, kori08}
and numerically \cite{kori08}. 

In this paper, by utilizing the above methodology by Kiss, Kori, Hudson,
and Rusin, we develop a general theory for the global feedback control
of  spatially extended oscillatory media.
Our approach is also based on a phase model.
Since the existence of diffusive coupling
plays a crucial role on the development of spatial patterns in
oscillatory media, our phase model inevitably includes both diffusive and global
coupling, in contrast to discrete oscillators.
Studying such a phase model, we find coupling functions leading to the
following spatial patterns characterized by the distribution of
phases: (i) 2-cluster states with specified population ratios, (ii)
equally populated multi-cluster states, and (iii) a desynchronized
state.
Moreover, we propose a new nonlinear feedback function without time
delay, which is more convenient to design various coupling functions
than that used in the previous work \cite{kiss07, kori08}.
We numerically demonstrate our proposed method by using a particular
reaction-diffusion model and reproduce all the above three patterns with
theoretically predicted feedback parameters.

This paper is organized as follows:
In section~\ref{sec:method}, we present the basic idea of
our control methodology for oscillatory media in detail. 
In section~\ref{sec:analysis}, we give a detailed analysis of the
phase diffusion equation with special coupling functions that yield the
above-mentioned three spatial patterns. 
Numerical demonstration of the theory by using the Brusselator is given
in section \ref{sec:brusselator}. 
Experimental realization is discussed in section~\ref{sec:discussion}.  

\section{General control methodology}
\label{sec:method}
Our approach to the control of oscillatory media is closely related to
the method recently proposed for the population of oscillators
\cite{kiss07,kori08}.  
Dynamics of discrete, identical limit-cycle oscillators under global feedback is
described by the following nonlinear dynamical equations:
\begin{eqnarray}
\frac{\rmd \bm u_i}{\rmd t}={\bm F}(\bm u_i) 
+\frac{K}{N}\bm e\sum_{j=1}^N h(\bm u_j),
\end{eqnarray}
where $\bm u_i$ is the state vector of the $i^{\rm th}$ oscillator 
($i=1, \cdots, N$), $\bm F$ is a nonlinear function describing a limit
cycle oscillation, $K$ is the coupling strength, 
$h(\bm u_i)$ represents the feedback, and $\bm e$ is a unit vector with
only one nonzero component: we have assumed that the feedback is
additively applied to the system.

When the coupling is weak, by treating the second term as a small
perturbation the system is reduced to
the phase model (\ref{eq:pm}) \cite{kuramoto84}. 
In this phase description (\ref{eq:pm}),
synchronization behavior depends solely on the coupling function
$\Gamma(\phi)$, and therefore one can control the synchronization behavior of
the system if the coupling function is freely given. 
It has been shown \cite{kiss07, kori08} that this can be done by applying a
properly designed external feedback signal $h(\bm u_i)$ to the
oscillators system. 
This method relies on the fact that the coupling function is the
convolution of the feedback $h$ and the phase response function $Z(\phi)$, 
which characterizes the sensitivity
of the phase to a weak external perturbation (see \ref{ap:derivation}). 

Since oscillatory media can be regarded as a population of oscillators
that are diffusively connected, 
we argue that the same method works for shaping the coupling function in the
phase description of oscillatory media. 
Consider a $d$-dimensional reaction-diffusion system with a global coupling:
\begin{equation}
\partial_t\bm{u}=\bm{F}(\bm{u})+ \hat{D}\nabla^2 \bm{u}
+\frac{K}{S}\bm e\int h(\bm{u})\rmd\bm{x}, 
\label{eq:rd}
\end{equation}
where $\bm{u}(\bm{x}, t)$ is the state vector, $\hat{D}$ denotes the
diffusion matrix, $\bm{F}(\bm{u})$ is 
a reaction term that generates a limit cycle oscillation with the frequency
$\omega$, $K$ is the coupling strength, $\bm e$ is the same as above, and
$h(\bm u)$ represents the feedback integrated over the 
entire space $S$. 
We assume that the system is Benjamin-Feir stable, \textit{i.e.}, 
the system undergoes spatially uniform oscillation when
$K=0$ (external control is absent). 
Following the standard procedure (see \ref{ap:derivation}), 
we obtain
\begin{eqnarray}
 \partial_t\phi (\bm{x},t)&=\omega + \alpha{
  \nabla}^2\phi+\beta({\nabla}\phi)^2 
+\frac{K}{S}\int\Gamma \bm( \phi(\bm{x})-\phi(\bm{x}') \bm) \rmd\bm{x}',
\label{eq:phase}
\end{eqnarray}
where $\phi(\bm{x},t)$ is
the phase of local oscillation; $\alpha$, and $\beta$ are constants
determined by the property of the oscillatory medium. 
As in the case of discrete oscillators, the
coupling function $\Gamma(\phi)$ can be arbitrarily shaped by
properly designing the feedback signal $h$. 

For discrete oscillators, $n$-cluster states can be generated from the
coupling function that contains $n^{\rm th}$ harmonics \cite{okuda93}. 
Even in the case of oscillatory media, the coupling function is expected
to work in the same way to stabilize the clustering pattern. A distinct
problem here, however, is that the spatial patterns are not solely 
determined from 
the global coupling but from the interplay between the diffusive
coupling and the global coupling, which makes the analysis much more
complicated.
For example, when clustering pattern forms, interfaces appear between the
clusters due to the diffusive coupling. 
Then controlling of the interface motion is required to obtain desired
clustering patterns.  

Hence we take the following strategy. 
We start from a coupling function with which discrete oscillators
described by equation (\ref{eq:pm}) exhibit clustering or desynchronization.
We then study the phase diffusion equation (4) with this coupling
function to find resulting spatial patterns.
Once the relation between the pattern and the coupling
function is obtained, the corresponding feedback function can be found
by following the same
procedure as the discrete oscillators case.
The key to carrying out this strategy is to find proper couping
functions that allows for analytical treatment of the phase equation.
Although analytical treatment is easy for a simple coupling function as
$\Gamma(\phi)=\sin(\phi+\theta)$ with a parameter $\theta$
\cite{mikhailov06}, only poor spatial patterns appear with such a
coupling function; higher harmonics in the coupling 
function are responsible for the formation of complex spatial patterns,
including phase clustering behavior.   
In the next section we propose such analytically tractable coupling
functions that produces
clustering states and the desynchronized state.

\section{Analysis of the phase model} 
\label{sec:analysis}
In this section, we study a one-dimensional phase diffusion equation
with a global coupling (\ref{eq:phase}). 
Here we propose coupling functions that yield interesting spatial
patterns; We are especially interested in two cluster states with specified
population ratios, equally populated multi-cluster states, and the
desynchronized state. 

\subsection{2-cluster states: numerical investigation}
Here we focus on the 2-cluster states with an arbitrary population
ratio. 
In particular, we look for well-defined 2-clusters, with the phases
maximally separated by $\pi$. 

Discrete oscillators are known to form various clustering states, the
behavior entirely governed by the form of the coupling function \cite{okuda93}. 
In special, the following coupling function yields 2-cluster states with
the two phase difference equal to $\pi$:
\begin{eqnarray}
 \Gamma(\phi)&=\sin\phi-\gamma\{\sin(2\phi+\theta)-\sin\theta\},
\label{eq:couple2}
\end{eqnarray}
where $\gamma$ and $\theta$ are parameters (see \ref{ap:2cl}). 
Equation (\ref{eq:pm}) with this coupling function has a family of
2-cluster solution with different population ratios of the two
clusters, which are stable for some range of population ratios. 
Thus, starting from a random initial condition the system converges to a
2-cluster state with its population ratio determined from the initial
condition.  

Using this coupling function, we numerically investigate the phase
diffusion equation (\ref{eq:phase}) in one dimension with the system size $S=L$, taking
$\theta$ as a control parameter and $\gamma=0.3$.
This equation is solved with the flux-free boundary condition by using the
second-order Euler scheme, with spatial and time interval being set to
$\Delta x=0.1$ and $\Delta t=0.01$, respectively. 
We set $L=100$, $K=0.1$, $\alpha=0.384\times 10^{-2}$. 
We assign several values to $\beta$ to make a phase diagram
below. Otherwise we set $\beta=1.089\times 10^{-2}$. This special choice
of $\alpha$ and 
$\beta$ is for later comparison to the Brusselator model.
\begin{figure}[t]
\begin{indented}
\item[]
  \includegraphics[keepaspectratio=t
 rue, width=12cm]{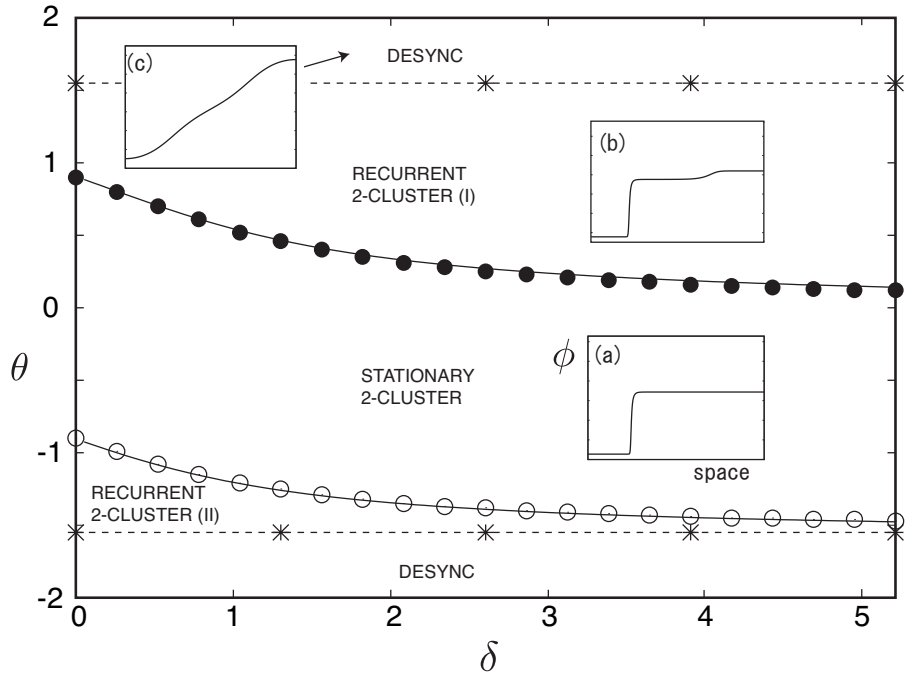}
\caption{\label{fig:phase}Phase diagram for 2-cluster states and the
 desynchronized state in the phase model with (\ref{eq:couple2}) for
 each $\delta=\beta/\alpha$.
 The points are numerical data, and the lines are given
 analytically. 
Recurrent 2-cluster has two ways of destabilization: the phase-advanced
 cluster becomes unstable (I), or the phase-retarded cluster becomes
 unstable (II).  
Inset figures are phase profiles with $\delta=2.83$ and (a) $\theta=0.23$, (b) $\theta=0.28$, (c)
 $\theta=1.58$. In (b) the pattern is not stationary; here a
 snapshot of budding a new cluster is shown.
}
\end{indented}
\end{figure}

\begin{figure}[t]
\begin{indented}
 \item[]
 \includegraphics[keepaspectratio=true, width=13cm]{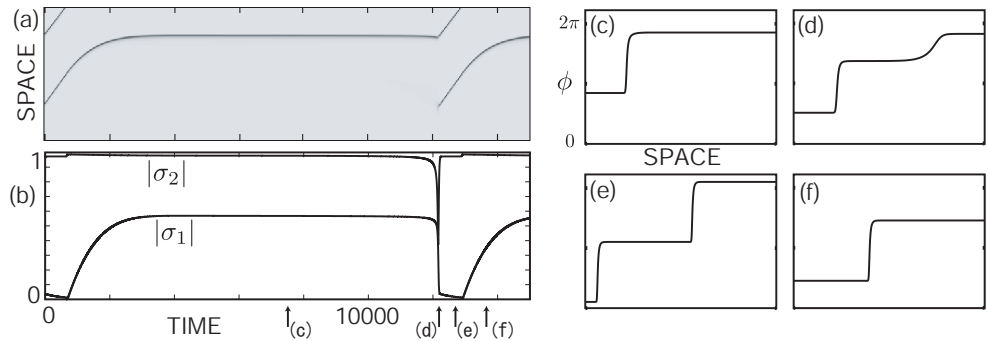}
\caption{\label{fig:spacetime} (a) Space-time plot of a recurrent
 2-cluster for $\alpha=0.384\times 10^{-2}$, $\beta=1.089\times 10^{-2}$ and $\theta=0.28$. 
Gray levels represent the spatial derivative of $\phi$, with the black lines
 indicating the location of the interfaces. 
(b) Corresponding time series of the amplitude of the order parameters
 $\sigma_1$ and  $\sigma_2$. 
(c)-(f) Snapshots of the phase profile, each corresponding to the arrows in (b). 
}
\end{indented}
\end{figure}

Figure~\ref{fig:phase} shows the phase diagram obtained by varying
$\theta$ for each $\delta\equiv\beta/\alpha$ with several values of
$\beta$ and fixed $\alpha$.  
As expected, there exists a finite range of stationary 2-cluster
states. 
Note that, as opposed to discrete oscillators, 
here the population ratio between the two clusters is uniquely determined for
fixed $\delta$ and $\theta$.
Increase (decrease) in $\theta$ widens the
phase-advanced (retarded) region. At some critical value of
$\theta$ the stationary state becomes unstable, leading to the
\textit{recurrent} 2-cluster state, where the following process occurs in a
repeated way [see figure~\ref{fig:spacetime}(a)]:
After a long transient of a quasi-stationary 2-cluster state, a new
cluster sprouts out of the phase-advanced (retarded) cluster.
Then the two interfaces propagate and one of the clusters disappear, the system
returning to the 2-cluster state.  
Such dynamics have been reported in CO oxidation model\
\cite{bertram03-the}, although investigated only numerically.

To characterize the patterns, we introduce the $l^{\rm th}$ order
parameters ($l=1, 2, \ldots$):
\begin{eqnarray}
 \sigma_l=\frac{1}{L}\int dx e^{-il\phi(x)}.
\end{eqnarray}
For 2-cluster states, $|\sigma_1|$ indicates an approximate
population disparity between the two clusters, and $1-|\sigma_2|$ the ratio of
the interface width to the system size $L$.
Note the two different
timescales in figure~\ref{fig:spacetime}(b), each corresponding to the
emergence of a new cluster and a slow drift of the interface.

As $\theta$ exceeds the threshold around $\pm \pi/2$, the
recurrent 2-clusters turn into the desynchronized state
[figure~\ref{fig:phase}(c)], where $\sigma_2$ almost vanishes. 

\subsection{2-cluster states: analytical investigation}
Here we analytically investigate the 2-cluster states numerically found
above. 
The analysis can be done by taking $L\to \infty$ limit. 
We derive analytical forms of $\sigma_1$ and $\sigma_2$
as functions of $\theta$ and give the stability boundaries of the stationary
2-cluster shown in figure~\ref{fig:phase}.
We move to a co-rotating frame so that the phase of the
phase-retarded cluster is fixed to $\phi=0$.  

As we can see from the numerical result, the profile of a 2-cluster state
can be decomposed into three regions: the phase-retarded cluster denoted by $A$ ($\phi=0$), the phase-advanced cluster denoted by
$B$, and the interface.  
Also, from the numerical observation it is implied that the instability leading to the {\it recurrent} 2-clusters appears from the clustered region, while the interface remains stable. 
Hence in the analysis below we assume that the interface does not
contribute to the stability.
This separation of the regions becomes well-defined for large $L$. 
When the interface width is negligible compared to the system size, the
two order parameters become real. 
In particular, in the steady state, we have $\sigma_2=1$,
so that $\sigma_1$ is the only relevant order parameter.  

Consider the dynamics of the cluster $A$.  
Contribution of the interface comes from the global coupling
represented as the integral in equation (\ref{eq:phase}). 
Since the interface width is vanishingly small, the interface region
itself does not affect the dynamics. 
The remaining effect of the interface comes indirectly through the
interface motion that varies the population ratio of the two clusters. 
However, since the timescale of the interface motion is $O(1/L)$, as
shown below, the population ratio can be treated as constant. 
Thus in this limit the dynamics of the clusters is independent of the
interface motion. 
When the interface can be negligible, equation (\ref{eq:phase}) has a
solution $\phi(x)=0$ for $x\in A$ and $\phi(x)=\pi$ for $x\in B$, where
the population ratio is arbitrarily given. 

The stability analysis can be performed in the same way as the discrete
oscillators (see \ref{ap:2cl}).  The only difference is the contribution
from the diffusive coupling, which turns out to be negligible in the
large $L$ limit.
Two modes of fluctuation occurs in the 2-cluster state: inter-cluster
and intra-cluster fluctuation. 
Inter-cluster mode is a fluctuation of the phase between the clusters, with each
cluster oscillating uniformly. The eigenvalue associated with this mode
is given by $\lambda_{\rm inter}=-1-2\gamma\cos\theta$. 
Thus by choosing $|\gamma|<\frac{1}{2}$ we can keep this mode stable. 
On the other hand, intra-cluster mode, a fluctuation within a cluster
can be unstable. 
The eigenvalues associated with the cluster $A$ and $B$ with the
wavenumber $k$ are given by 
$\lambda_{\rm intra}^{(A)}=-\alpha k^2 + 2p-1-2\gamma\cos\theta$ and 
$\lambda_{\rm intra}^{(B)}=-\alpha k^2 + 1-2p-2\gamma\cos\theta$,
respectively, where $p$, the population ratio, is the area fraction of
the cluster $A$ and is related to $\sigma_1$ through $2p-1=\sigma_1$. 
The negative sign of $k^2$-terms implies that the diffusion always works
as stabilizing the ${\rm inter}$-cluster modes; the most
unstable mode is the one with the smallest (but finite) wavenumber. 
Taking the large $L$ limit, this smallest wavenumber is vanishingly
small, so that the $k^2$-terms can be dropped from the expression of the
eigenvalue. 

Thus the diffusion does not affect the stability, while the stability
depends on the population ratio.  
To obtain the analytical expression of the population ratio, let us
consider the interface dynamics.
The two clusters $A$ and $B$ are treated as the fixed boundaries of the
interface. 
Since the inter-cluster mode is stable, the boundary conditions of the
interface profile are given by 
$\phi(-\infty)=0$ and $\phi(\infty)=\pi$, and $\sigma_2$ is replaced by
the steady-state value, $\sigma_2=1$.  
Then equation (\ref{eq:phase}) becomes
\begin{eqnarray}
  \partial_t\phi&=&\partial_x^2\phi+\delta(\partial_x\phi)^2
+\sigma\sin\phi-\gamma(\sin(2\phi+\theta)-\sin\theta), 
\label{eq:phase2}
\end{eqnarray}
where we have defined $\sigma\equiv \sigma_1$ and $\delta\equiv
\beta/\alpha$, and rescaled time and space as $Kt\to t$ and
$\sqrt{K/\alpha}\ x \to x$. 

Suppose that there is a traveling solution $\phi=f(x-ct)$, where the
interface velocity $c$ is written as $c=\frac{L}{2}\frac{d\sigma}{dt}$,
owing to the fact that interface motion varies the population ratio.   
Multiplying (\ref{eq:phase2}) by $\partial_xf$ and integrating
over the entire space yields 
\begin{eqnarray}
  \frac{d\sigma}{dt}&=-\frac{4}{L\int (\partial_xf)^2dx}\left\{
\sigma + \frac{\delta}{2}\int (\partial_xf)^3dx
+\frac{\pi\gamma\sin\theta}{2}
\right\}.
\label{eq:dsigmadt}
\end{eqnarray}
Therefore $\sigma$ has a stable solution formally written as
\begin{eqnarray}
\sigma=-\frac{\delta}{2}\int(\partial_xf)^3dx
-\frac{\pi\gamma\sin\theta}{2}.\label{eq:sig_stationary} 
\end{eqnarray}
From (\ref{eq:dsigmadt}) it is seen that the interface slowly moves with the timescale of $O(1/L)$
toward this stable state. 
Thus the interface dynamics, or the time evolution of $\sigma$, is
decoupled from the rest. 

An explicit expression of the steady state solution of $\sigma$ can be
obtained through perturbation expansion.  
First we seek for a stationary solution of
(\ref{eq:phase2}).  When $\theta$ satisfies $\tan\theta=-\delta$,
there exists an exact solution connecting $\phi(-\infty)=0$ and
$\phi(\infty)=\pi$:
\begin{eqnarray}
\phi_0(x)=2\arctan e^{\kappa x},
\end{eqnarray}
where $\kappa=\sqrt{2\gamma\cos\theta}$.
It is easily verified from (\ref{eq:sig_stationary})
that this solution satisfies $\sigma=0$. Then perturbation expansion can
be performed in terms of $\sigma$ up to the first order. 
We obtain (see \ref{ap:sigma}):
\begin{eqnarray}
 \sigma&=-\frac{2\gamma\cos\theta}{\chi(\delta)}
\left(\delta + \tan\theta\right),
\label{eq:sigma}
\end{eqnarray}
where
$\chi(\delta)=\frac{2\delta(1+\delta^2)}{1+4\delta^2}\coth\pi\delta$.

\begin{figure}[t]
\begin{indented}
 \item[]
  \includegraphics[keepaspectratio=true, width=8cm]{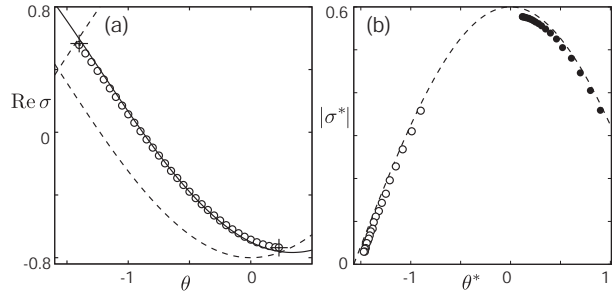}
\caption{\label{fig:sgm}(a)Relation between the real part of $\sigma$ and
 $\theta$ for $\alpha=0.384\times 10^{-2}$ and $\beta=1.089\times 10^{-2}$. 
The open circles are numerical data;
the dotted lines represent the boundary of the existence of stationary 2-cluster states, given by (\ref{eq:st_cond});
the solid line is given by (\ref{eq:sigma});
crosses at $\theta=-1.40$ and $\theta=0.23$ are numerical data from the
 Brusselator model with corresponding parameters.  
(b) Relation between critical values of $\sigma$ and $\theta$, denoted
 by $\sigma^*$  and $\theta^*$. The filled and open circles correspond to
 the instabilities of phase-advanced and phase-retarded clusters,
 respectively, with different values of $\beta$. 
 The dotted line is  (\ref{eq:st_cond}). }
\end{indented}
\end{figure}

Substituting the expression of $\sigma$ into the intra-cluster
eigenvalues, we obtain the stability condition of 2-cluster states. 
In the large $L$ limit, the $k^2$-term in the eigenvalues vanishes and
the stability boundary is given by 
\begin{eqnarray}
 \sigma\pm 2\gamma\cos\theta=0.\label{eq:st_cond}
\end{eqnarray}
Figure~\ref{fig:sgm} shows the dependence of $\sigma$ on $\theta$. 
We have plotted only the real part of $\sigma$, while in our
simulation the imaginary part is $O(K)$ and is negligible. 
Both equations (\ref{eq:sigma}) and (\ref{eq:st_cond}) fit well with the
numerical data.  
Moreover, substituting (\ref{eq:sigma}) into (\ref{eq:st_cond})  yields
\begin{eqnarray}
 \tan\theta=-\delta\mp\chi(\delta),
\label{eq:pd_cond}
\end{eqnarray}
which gives the threshold value of $\theta$ for the stability of stationary
2-cluster states. 
Note that the stability condition is independent of $\gamma$. 
The theoretical lines given by (\ref{eq:pd_cond}) are in excellent
agreement with numerical data in figure~\ref{fig:phase}. 

Thus, within the range of parameter $\theta$ determined from
(\ref{eq:pd_cond}), we can control $\sigma$ as a function of
$\theta$ via (\ref{eq:sigma}). 

Now the interpretation of the recurrent 2-cluster states are given as
follows: given an initial condition, the system converges to a 2-cluster
state with a slowly-moving interface. When $\sigma$, varies through the
interface motion, exceeds the threshold given by equation
(\ref{eq:st_cond}), the intra-cluster mode becomes unstable and one of
the clusters collapses. 
Since the inter-cluster mode remains stable, the system returns to a
2-cluster state with reduced $\sigma$, the whole process repeated
\textit{ad infinitum}.  

\subsection{Desynchronized state}
We give a theoretical analysis of equation (\ref{eq:phase2}) for the
desynchronized state.  
The perfect desynchronized state is defined such that all the order
parameters vanish. 
However, in practice some order parameters remain finite because of the
boundary effect for flux-free boundary conditions. 
 
Firstly, as an ideal case, let us assume the periodic boundary condition. 
Then the perfect desynchronized state can be given by $\phi(x)=2\pi x/L$.  
Linear stability analysis for this profile shows that for each mode
with the wavenumber $k_l=2\pi l/L$ $(l\ge 1)$ the corresponding
eigenvalue is 
$\lambda_{\mathrm{desync}}^{(l)}=\lambda_0^{(l)}-\alpha k_l^2-2\beta i
k_lk_1$, where $\lambda_0^{(1)}=-\frac{1}{2}$, $\lambda_0^{(2)}=\gamma
e^{i\theta}$, and $\lambda_0^{(l\ge 3)}=0$.  
Hence the $l=2$ mode loses the stability at $\theta=\pm \frac{\pi}{2}$ in
the large $L$ limit, which fits well with numerical data in
figure~\ref{fig:phase}. 
Note that, if the diffusive coupling is absent, only $l=1$ and $l=2$
modes are stable, $l\ge 3$ modes being neutral. 

Since the boundary is not periodic but flux-free in the present case,
the profile deviates from the linear one, as shown in
figure~\ref{fig:phase}(c),
Accordingly, the steady state values of $\sigma_l$
are shifted from zero by $O(1/L)$ (order of the width of the
boundaries).
Note that in the linear regime the main contribution to $\sigma_l$ comes
from the mode $k_l$. 
The modes $l=1$ and $l=2$ have the eigenvalues of order $O(1)$ as seen
above, and thus $\sigma_1$ and $\sigma_2$ remains to be $O(1/L)$. 
On the other hand, since $l\ge 3$ modes have the eigenvalues of
$O(1/L^2)$, nonlinear effects of order $O(1/L^2)$ coming from the
terms such as $\sigma_1\sigma_2$ makes $l\ge 3$ modes grow up to $O(1)$. 
Therefore, in order to get better desynchronized state, we need to add
as many higher harmonics as possible, as demonstrated in section \ref{sec:brusselator}. 

\subsection{Multi-cluster states}
The arguments of 2-cluster states can be extended to $n$-clusters in the following way. 
Consider the following coupling function:
\begin{eqnarray}
 \Gamma(\phi)=\sum_{m=1}^{n-1}\sin m\phi
-\gamma\{\sin (n\phi+\theta)-\sin\theta\}.
\label{eq:gamma}
\end{eqnarray}
This coupling function, when introduced to discrete oscillators,
creates stable equally-populated $n$-clusters with the phases evenly
separated (\ref{ap:2cl}). 
Let us find a stationary, equally populated $n$-cluster solution of
(\ref{eq:phase}) with (\ref{eq:gamma}). Such a solution
satisfies $\sigma_m=0$ ($m<n$) and $\sigma_n=1$, and hence only $n^{\rm th}$
harmonic remains in (\ref{eq:phase}).  
Then, by choosing $\theta$ so as to satisfy
\begin{eqnarray}
\delta=-\frac{n}{2}\tan\theta, \label{eq:theta_n}
\end{eqnarray}
we have a solution with each cluster separated by $\frac{2\pi}{n}$
and all the $n-1$ interfaces having the same interface profile given by
$\phi(x)=\frac{4}{n}\arctan\exp(\kappa_n x)$ with
$\kappa_n=\sqrt{n\gamma\cos\theta}$. 
This state is stable against {\it inter-} and {\it intra-}cluster
fluctuations (see \ref{ap:2cl}).

For the stability of the desynchronized state, the same argument as in 
the above $n=2$ case holds and the stability boundary is given by
$\theta=\pm \pi/2$. 

\section{Numerical confirmation with the Brusselator model} 
\label{sec:brusselator}
The above analytical expressions are used for the control of oscillatory
media. 
As a model system of oscillatory media, we adopt the Brusselator model: 
\begin{eqnarray}
\frac{\partial u}{\partial t}=D_u\nabla^2u + A-(B+1)u+u^2v
+\frac{K}{L^d}\int h(u,v) \rmd\bm x, \label{eq:bru_u}\\ 
\frac{\partial v}{\partial t}=D_v\nabla^2v +Bu-u^2v.\label{eq:bru_v}
\end{eqnarray} 
The parameters $A$, $B$, $D_u$, and $D_v$ are chosen in such a way that
the system exhibits stable uniform oscillation;
we set $A=1.6$, $B=5.0$, $D_u=0.01$, and $D_v=0$. 
The corresponding parameters in the phase model are
$\alpha=0.384\times 10^{-2}$ and $\beta=1.089\times 10^{-2}$. 
For the precision that assures the validity of the phase description, we set $K=0.001$ and $L=1000$
(equivalent to $K=0.1$ and $L=100$ in the phase model).  
Note that while for convenience of numerical simulation we have set
$D_v=0$, we may also consider nonzero $D_v$, which simply results in
the variation in the values of $\alpha$ and $\beta$. 

As a feedback function we propose the following:
\begin{equation}
 h(u,v)=h(\phi)= \sum_{n=0}^M k_n\cos \bm{(}n\phi(u,v) - \psi_n\bm{)},
  \label{eq:h}
\end{equation}
where $\phi(u,v)$ is the phase of the limit cycle oscillation
\footnote{In our simulation, the phase $\phi(u,v)$ can be obtained directly from
$u$ and $v$ in the following way: We first define the phase on the
unperturbed ($K=0$) 
limit cycle so that the phase evolves with a constant velocity, which
can be done numerically. 
We then define the phase of a point$(u,v)$ off the limit cycle by the phase of the nearest point on the limit cycle. 
}, and the
parameters $k_n$ and $\psi_n$ are the feedback intensity and the 
phase shift of the $n^{\rm th}$ feedback term, respectively. 
The coupling function is obtained from $h(\phi)$ and the phase response
function $Z(\phi)$, which characterizes the sensitivity of the phase to
a weak external perturbation (see \ref{ap:derivation}).  
By expanding $Z(\phi)=\sum_{l}z_l\cos(n\phi+\chi_l)$, the coupling function is
written as 
\begin{eqnarray}
 \Gamma(\phi)&=\sum_{l=0}^M\frac{z_l k_l}{2}\cos(l\phi+\psi_l+\chi_l).
\label{eq:gam_four}
\end{eqnarray}
In principle, as long as $z_l$ is finite, we can
assign any value to $l^{\rm th}$ harmonics of the coupling function by
choosing appropriate values for $k_l$ and $\psi_l$. 
The advantage of using (\ref{eq:h}) is that the relation between
parameters in the coupling function and the feedback parameters
${k_n,\psi_n}$ is given in a simple manner. 
(We could also use as the feedback $h(u, v)$ a polynomial of $u$ with multiple time delays
\cite{kiss07, kori08}, but in that case the relation is
represented as a nonlinear function and the parameters need to be
calculated numerically.)
Hence, given a coupling function, we can calculate the corresponding
feedback parameters by measuring phase response function $Z(\phi)$. 
Table \ref{tab:params} shows $Z(\phi)$ and the feedback parameters 
corresponding to (\ref{eq:gamma}) for $n=2$ and $n=5$. 
In the following numerical investigation we set $\gamma=0.3$. 

First we study 2-cluster states in the one-dimensional case with the
feedback parameters corresponding to $n=2$. 
We have confirmed that, for several parameter values of $\theta$ we can
observe stationary 2-clusters, recurrent 2-clusters, and desynchronized
states, with the order parameter values predicted by the phase model
(deviation of order $O(10^{-2})$).  
As an example, 
in figure~\ref{fig:sgm}(a), numerically obtained critical values of the
real part of $\sigma_1$ (denoted by $''+''$) are superimposed on the
data from the phase model, which are in good agreement with the
corresponding phase model, with deviations $O(10^{-2})$. 

\begin{figure}[t]
\begin{indented}
 \item[]
 \includegraphics[keepaspectratio=true, width=8cm]{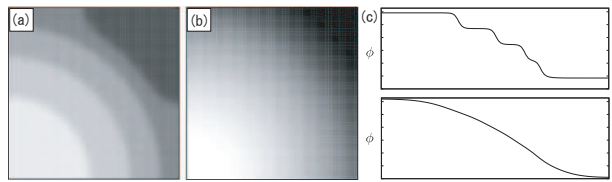}
\caption{\label{fig:2d}Two dimensional stationary pattern (uniform
 rotation subtracted) of the Brusselator model with global feedback
 containing 5 harmonics: (a) nearly equally  populated 5-cluster
 ($\theta=-0.848$) and  (b) desynchronized state  ($\theta=-1.580$). 
(c) Section plots of (a)(top) and (b)(bottom), each from bottom left to top right. }
\end{indented}
\end{figure}

Next, we use (\ref{eq:gamma}) for $n=5$ to produce the equally
populated 5-cluster state and the desynchronized state in the
two-dimensional case.  
The 5-cluster is shown in figure~\ref{fig:2d}(a), the parameter $\theta$
given by (\ref{eq:theta_n}) with $n=5$. 
The order parameters are $|\sigma_5|=0.739$ and $|\sigma_l|\sim O(10^{-2})$ for
$l<5$, indicating that the five clusters are well-defined and
approximately equally populated. 
In figure~\ref{fig:2d}(b) we have a desynchronized state with 
$\theta$ just below the threshold ($\theta=-1.58$), where
$|\sigma_l|\sim O(10^{-3})$ for $l\le 5$ and $O(10^{-2})$ for $l>5$. 
Note that the degree of desynchronization becomes better than the one
for $n=2$ shown in section~\ref{sec:analysis}; we can make a better
desynchronized state by adding appropriate higher harmonics.  

\begin{table}
\caption{\label{tab:params}Numerically obtained response function
 $Z(\phi)=\sum_{l}z_l\cos(l\phi+\chi_l)$ for the Brusselator with
 $A=1.6$ and $B=5.0$, and the feedback  parameters $\{k_l\}$ and
 $\{\psi_l\}$ in (\ref{eq:h}) producing the coupling functions
 given by (\ref{eq:gamma}) with $n=2$ and $n=5$. For $l>n$, $k_l$ and
 $\psi_l$ are equal to zero.}
\begin{indented}
\lineup
\item[]
\begin{tabular}{cllllll}
\br
$l$ & \centre{1}{$z_l$} & \centre{1}{$\chi_l$} & 
\centre{1}{$k_l$} & \centre{1}{$\psi_l$} & 
\centre{1}{$k_l$} & \centre{1}{$\psi_l$}  \cr
 & & & \centre{2}{(for $n=2$)} & \centre{2}{(for $n=5$)}\cr
\mr
0 &  0.8618 & 0.0 & \m2.320$\gamma$ & $-1.570+\theta$ 
& \m\0$2.320\gamma$ & $-1.570+\theta$\cr
1 & 1.792 & 1.174 & \m1.115 & $-$2.745 & 
\m\01.115 & $-$2.745 \cr
2 & 0.6390 & 1.410 & $-3.129\gamma$ & $-2.981+\theta$
& \m\03.129 & $-$2.981 \cr
3 & 0.3701 & 2.441 &  &  & \m\05.403 & $-$4.012\cr
4 & 0.1696 & 2.037 &  &  &\m11.78 & $-$3.608\cr
5 & 0.03714 & 1.595 &  &  & $-$53.84$\gamma$ &$-3.165+\theta$\cr
\br
\end{tabular}
\end{indented}
\end{table}

\section{Discussion: experimental realization of the theory}\label{sec:discussion}
To apply our method to experimental systems, we need to find the
constants $\alpha$ and $\beta$, and the response function $Z(\phi)$. 
Since it is generally expected that the target pattern appears in the
oscillatory media (due to inhomogeneities, or by applying a manual
stimulus) \cite{kuramoto84}, $\beta$ can be measured by using the target pattern, assuming its phase profile as $\phi(\bm x,t)=\Omega t+k|\bm x|$ (the
origin is on the center of the target pattern) with the 
measurable quantities $k$ and $\Omega$: 
Substituting this expression into (\ref{eq:phase}) with $K=0$ we obtain
$\beta=(\Omega-\omega)/k^2$. 
The decay rate of the local perturbation from a
uniform oscillation gives $\alpha$. 
The response function $Z(\phi)$ can
be measured by perturbing the system through a global parameter, to
which we also apply the global feedback. 
To use (\ref{eq:h}), instantaneous measurement of the phase
$\phi(\bm{x},t)$ at each spatial point is needed. 
If at least one quantity of an oscillator is observable, this can be
done by, for example, constructing a delayed coordinate.  

Moreover, the following things should be taken into account for experimental
realization of our theory. 
First, feedback must be weak for the precision of the phase
description. This implies that the system size should be large enough to
obtain well-defined cluster states even under weak feedback: the width
of the interface is $O(\sqrt{D/K})$, which must be sufficiently smaller
than the linear dimension $L$. Also, to make a coupling function
containing large enough higher harmonics with weak feedback, oscillation
is better to be relaxation type: then the response function has higher
harmonics with large amplitudes, so that we can keep the feedback signal
weak to realize a desired coupling function [see the expression of
$\Gamma(\phi)$ (\ref{eq:gam_four})]. Second, the emergence of
phase singularity leads to the breakdown of the phase description and
must be avoided. 

We have checked in our preliminary numerical simulations that the
multi-cluster states and the desynchronized state are robust against
noise. 
Thus we are convinced that our proposed method works in experimental systems.

\section{Concluding remarks}
We have proposed a theoretical framework for designing spatial patterns in
oscillatory media. 
When a certain pattern is found in a phase model with a specific
coupling function, the same pattern can be realized in oscillatory media by
applying a properly constructed nonlinear feedback. 
In this paper, we found analytically tractable coupling functions that
enables us to quantitatively control the spatial patterns. 
Using these coupling functions, we investigated the phase equation with
the global coupling and found the parameter regions where the following
patterns stably exist: 2-cluster states with specified population ratios,
equally populated multi-cluster states, and the desynchronized state. 
In the case of 2-clusters, we gave analytical expression of the population
ratio of the two clusters as the function of a feedback parameter.  
We also proposed a simple form of the nonlinear feedback function to make the
calculation of the feedback parameters easier. 
We exemplified all these results using the Brusselator model and succeeded to reproduce
the patterns predicted by the phase model. 
Since our method is based on the measurable quantities only, it is
expected that the method is verified in a real experiment. 

The desynchronized state deserves further remark. 
Our results show that even in oscillatory media one can drive the
system into the desynchronized state, as well as in discrete oscillators \cite{kiss07}. 
Such a control is not only of medical \cite{kiss07, ottino07}, but also
potentially of industrial interest; for example, it would be
beneficial when constant output from oscillatory catalytic
reaction is desirable.   

Further investigation of the phase model with other coupling
functions is of great interest for controlling more complex patterns,
although our method is limited to oscillatory system and cannot be
applied to some typical spatial patterns such as the Turing pattern. 
Also, investigating the control of Benjamin-Feir unstable systems by
replacing the phase diffusion equation with Kuramoto-Sivashinsky
equation will be interesting both in a theoretical sense
and for application.

\ack
The authors are grateful to Y. Nishiura and A.~S. Mikhailov for valuable
discussions.

\appendix
\section{Derivation of the phase model}
\label{ap:derivation}
In this Appendix we derive the phase model
(\ref{eq:phase}) from a reaction-diffusion system with a global
feedback. 
The system is assumed to undergo spatially uniform oscillation when
external control is absent (namely, the system is Benjamin-Feir stable
\cite{kuramoto84}). 
Dynamical evolution of a $d$-dimensional oscillatory medium is
described by a reaction-diffusion equation:
\begin{equation}
\partial_t\bm{u}= \bm{F}(\bm{u};q) + \hat{D} \nabla^2 \bm{u}. 
\label{eq:rd_original}
\end{equation}
Note that here we consider a general situation, where the global feedback is
introduced through a global parameter $q$. 
In equation (\ref{eq:rd}), and in References \cite{kiss07, kori08}, the
feedback is simply applied additively. 
External feedback is
applied to $q$ as
\begin{equation}
 q(t) = q_0 + K p(t),
\end{equation}
where $q_0$ and $K>0$ are constants. By assumption, $\partial_t\bm{u} =
\bm{F}(\bm{u};q_0)$ yields a limit-cycle oscillation, with its solution
denoted by $\bm u=\bm u_0(t)$. The function
$p(t)$ describes a global feedback signal, given by
\begin{equation}
 p(t) = \frac{1}{S} \int h(\bm u) \rmd \bm{x},
\end{equation}
where $h(\bm u)$ is some feedback function. 
The integration is taken over the entire space and $S$ is the
volume of the system.
(Various functions can be considered for $h$. Our particular choice has
been given in equation \ref{eq:h}. )

As we have assumed, feedback intensity $K$ is small, so that by dropping
$O(K^2)$ equation (\ref{eq:rd_original}) can be approximated by
\begin{equation}
\partial_t\bm{u}=\bm{F}(\bm{u};q_0)+ \hat{D}\nabla^2 \bm{u}+K
 p(t) \bm{f}(\bm{u}),  
\label{eq:rd_generalized}
\end{equation}
where $\bm{f}(\bm{u}) \equiv (\partial \bm {F}/ \partial q)_{q=q_0}$.
When $\bm f$ is independent of $\bm u$, the global parameter $q$ appears
additively and the system reduces to equation 
(\ref{eq:rd}). 

When a spatial pattern emerges for small $K>0$, the spatial
variation, and thus $\nabla^2\bm u$, is expected to be small, vanishing
as $K\to 0$. Thus, in addition to the feedback term, we may treat the
diffusion term as small perturbations to the 
limit cycle (this is the case in our simulation, where the interface
width is $O(\sqrt{D/K})$), and therefore the diffusion term is the same
order as the feedback).
Then, following a standard method developed by
Kuramoto \cite{kuramoto84}, we can derive a closed description for the
phase variable for our oscillatory medium. 
As is usually adopted, the phase $\phi(\bm u)$ is defined so as to satisfy $\partial_{\bm u}\phi\cdot\bm
F(\bm u; q_0)=\omega$. 
Substituting this relation into the identity 
$\partial_t\phi=\partial_{\bm u}\phi\cdot\partial_t{\bm u}$,
we obtain
\begin{eqnarray}
\partial_t\phi=\omega+\partial_{\bm u}\phi\cdot\left\{
\hat{D}\nabla^2 \bm{u}+K p(t) \bm{f}(\bm{u})
\right\}.\label{eq:phi_nonav}
\end{eqnarray}
At the lowest order of $K$, we can replace $\bm u$ with the
value on the limit cycle $\bm u_0$. 
Then the equation above is expressed only in terms of $\phi$. 
After averaging (\ref{eq:phi_nonav}) over one period of oscillation, we
arrive at equation (\ref{eq:phase}), 
where $\alpha$, $\beta$, and $\Gamma(\phi)$ are written as
\begin{eqnarray}
 \alpha = \frac{1}{2\pi}\int_0^{2\pi} \rmd\phi
\bm {\tilde Z}(\phi)\hat{D}\frac{\partial \bm u}{\partial \phi}, 
\label{eq:alpha} \\
 \beta = \frac{1}{2\pi}\int_0^{2\pi} \rmd\phi
\bm {\tilde Z}(\phi)\hat{D}\frac{\partial^2 \bm u}{\partial \phi^2} , 
\label{eq:beta} \\
 \Gamma(\phi-\phi') =\frac{1}{2\pi} \int_0^{2\pi} \rmd\lambda
  Z(\phi+\lambda) h(\phi'+\lambda). \label{eq:gam_resp}
\end{eqnarray}
Here, the phase response function $Z(\phi) \equiv \bm{\tilde Z}(\phi)
\cdot \bm{f}(\phi)$, defined as the response to the global parameter
$q$, and the ``bare'' response function $\bm{\tilde Z}(\phi) \equiv
\partial_{\bm u} \phi|_{\bm u=\bm u_0}$, are evaluated on the
unperturbed limit-cycle orbit.  

Expanding $Z(\phi)$ and $h(\phi)$ as $Z(\phi)=\sum_{l=-\infty}^{\infty}
z_le^{il\phi}$ and $h(\phi)=\sum_{l=-M}^M h_le^{il\phi}$ respectively,
we obtain 
\begin{equation}
 \Gamma(\phi)=\sum_{l=-M}^M z_lh_{-l}e^{il\phi}. \label{eq:gam_conv}
\end{equation}
Hence the coupling function containing up
to the $M^{\rm th}$ harmonics can be generated by determining $h(\phi)$
up to the $M^{\rm th}$ harmonics, as long as $z_l$ has a finite
value\cite{kiss07, kori08}. 

\section{2-cluster states for the coupled oscillators}
\label{ap:2cl}Here we show that the collection of discrete oscillators interacting through the coupling
function given by (\ref{eq:couple2}) can exhibit 2-cluster states
with their phases separated by $\pi$.  

\subsection{Steady-state 2-cluster solution}
Consider a set of $N$ identical oscillators with the frequency
$\omega$. 
The dynamics is written as
\begin{eqnarray}
\dot{\phi_i}=\omega+\frac{K}{N}\sum_{j=1}^N\Gamma(\phi_i-\phi_j). 
\end{eqnarray}
To produce $n$-cluster states, it is sufficient that the coupling function
contains up to the $n^{\rm th}$ harmonics \cite{okuda93}; linear
stability analysis shows that the harmonics smaller than n does not
contribute to the stability of $n$-cluster states, and the $n^{\rm th}$ harmonics works in a similar
way to $n$:1 periodic forcing \cite{mikhailov06}. 
In special, to observe 2-cluster states, one needs to prepare the
coupling function such that the first harmonics destabilizes the
1-cluster, \textit{i.e.}, perfect synchronization, and the second
assures the 2-cluster. 
Thus the coupling function for 2-cluster states can be written as
\begin{eqnarray}
 \Gamma(\phi)&=\sin(\phi+\theta_1)-\gamma\sin(2\phi+\theta_2).
\end{eqnarray}
The amplitude of the first harmonics can be absorbed into the coupling
constant $K$. Also, for later convenience we choose the negative sign
for the second harmonics.

Assume that the oscillators form a 2-cluster state, where $N_A$ 
oscillators belong to the cluster $A$ with the phase $\phi_A$ and
$N-N_A$ to the cluster $B$ with $\phi_B$. 
In the phase-locking state ($\dot\phi_i=\Omega$ for all $i$), we get
\begin{eqnarray}
 \Omega&=p\Gamma(0)+(1-p)\Gamma(\psi), \\
 \Omega&=p\Gamma(-\psi)+p\Gamma(0), 
\end{eqnarray}
where $\psi=\phi_A-\phi_B$, $p=N_A/N$ and $\Omega=\omega+\sin\theta_1-\gamma\sin\theta_2$ is
the frequency of the clusters. 
Then $\psi$ satisfies
\begin{eqnarray}
(2p-1)\Gamma(0)+(1-p)\Gamma(\psi)-p\Gamma(-\psi)=0.  \label{eq:app_psi}
\end{eqnarray}
When we choose $\theta_1=0$, (\ref{eq:app_psi}) has a solution $\psi=\pi$
for any $p$.  
In special, when $|\gamma|<\frac{1}{2}$, $\psi=\pi$ is the only solution 
except for $\psi=0$, the single cluster solution. 
If $\theta_1\neq 0$, the phase difference is shifted from $\pi$ except
for $p=\frac{1}{2}$. 

\subsection{Linear stability analysis}
We perform the linear stability analysis by expanding $\phi_j$ as 
$\phi_j=\phi_j^{(0)}+\xi_j$, where $\phi_j^{(0)}=0$ for $j\in A$ and
$\phi_j^{(0)}=\pi$ for $j\in B$. 

First we consider the \textit{inter}-cluster mode, where the fluctuation
is uniform in each cluster. 
In this case we can write $\xi_{j\in A}=\xi_A$ and $\xi_{j\in B}=\xi_B$. 
The mode $\xi_A$ obeys
\begin{eqnarray}
 \dot{\xi_A}&=K(1-p)\Gamma'(\pi)(\xi_A-\xi_B), \\
 \dot{\xi_B}&=Kp\Gamma'(-\pi)(\xi_B-\xi_A).
\end{eqnarray}
Changing the variables as $\xi_{\pm}=\xi_A\pm \xi_B$, and using
$\Gamma'(\pi)=\Gamma'(-\pi)=-1-2\gamma\cos\theta$, we obtain
\begin{eqnarray}
\dot{\xi}_+&=K(1-2p)(1+2\gamma\cos\theta)\xi_{-}, \\
 \dot{\xi}_-&=-K(1+2\gamma\cos\theta)\xi_{-}.
\end{eqnarray}
The zero mode $\xi_+$ represents uniform rotation along with the limit
cycle. On the other hand, $\xi_-$ corresponds to the inter-cluster mode,
which is stable regardless of $\theta$ for $|\gamma|<\frac{1}{2}$. 

Next we consider the \textit{intra}-cluster mode, where the fluctuation
occurs within each cluster and the spatial average of $\xi$ within each
cluster is zero. We get
\begin{eqnarray}
 \dot{\xi}_{j\in A}=&K(2p-1-2\gamma\cos\theta)\xi_{j\in A}, \\
 \dot{\xi}_{j\in B}=&K(1-2p-2\gamma\cos\theta)\xi_{j\in B}.\\
\end{eqnarray}
Thus the intra-cluster fluctuation the eigenvalues
$\lambda_A=K(2p-1-2\gamma\cos\theta)$ and
$\lambda_B=K(1-2p-2\gamma\cos\theta)$.  
These modes can be destabilized depending on the population ratio $p$. 
In special, when $|\theta|>\pi/2$, either $\lambda_A$ or $\lambda_B$ is
positive for any $p$. 

Similarly, the coupling function (\ref{eq:gamma})
produces $n$-cluster solutions $\phi_i=2\pi l/n$ ($l=0, \cdots, n-1$)
for arbitrary population ratio, as can be checked by direct substitution. 
Stability is studied analogously with the $n=2$ case above, the {\it
intra}- and {\it inter}-cluster eigenvalues given by 
\begin{eqnarray}
\lambda_{\mathrm{inter}}^{(n)}&=-K\left(\frac{n}{2}+n\gamma\cos\theta\right), \\
 \lambda_{\mathrm{intra}}^{(n)}&=-K\left(\frac{n}{2}-\frac{n^2}{2}p_m+n\gamma\cos\theta\right),
\end{eqnarray}
where $p_m$ is the fraction of the $m^{\rm th}$ cluster. 
When the clusters are equally populated,
$\lambda_{\mathrm{intra}}^{(n)}=-Kn\gamma\cos\theta$ and is stable for
$|\theta|<\pi/2$. Conversely, when $|\theta|>\pi/2$ at least one of $n$
clusters has positive $\lambda_{\mathrm{intra}}^{(n)}$ and the $n$-cluster
state is no longer stable. 

\section{Derivation of (\ref{eq:sigma})}\label{ap:sigma}
We expand the parameter $\theta$ and the profile $\phi(x)$ in terms of
$\sigma$ as follows:
\begin{eqnarray}
\tan\theta=-\delta+\sigma\mu_1 + O(\sigma^2), \\
\phi(x)=\phi_0(x)+\sigma\phi_1(x) + O(\sigma^2).
\end{eqnarray}
Substituting these expressions into (\ref{eq:phase2}) yields the
following linearized equation:
\begin{eqnarray}
 \mathcal{L}&\phi_1(x)=
-2\gamma\cos\theta\sin\phi_0(x)-\mu_1\sin^2\phi_0(x), 
\label{eq:solv_cond}
\end{eqnarray}
where the linearized operator $\mathcal{L}$ is given by
\begin{eqnarray}
\mathcal{L}=&\kappa^{-2}\partial_x^2-\cos 2\phi_0(x)
+2\delta\sin\phi_0(x)\bm{(}\kappa^{-1}\partial_x-\cos\phi_0(x)\bm{)}.
\end{eqnarray}
The adjoint operator $\mathcal{L}^{\dag}$ is written as 
\begin{eqnarray}
\mathcal{L}^{\dag}=&\kappa^{-2}\partial_x^2-\cos 2\phi_0(x)
-2\delta\sin\phi_0(x)\bm{(}\kappa^{-1}\partial_x+2\cos\phi_0(x)\bm{)}.
\end{eqnarray}

It is verified by direct calculation that $\mathcal{L}^{\dag}$  has the
zero eigenfunction $\exp[2\delta\phi_0(x)]\mathrm{sech}\kappa x$. 
The solvability condition reads
\begin{eqnarray}
 \int_{-\infty}^{\infty}\!\!\!dx\ \exp[2\delta\phi_0(x)]\mathrm{sech}\kappa x
\left(
-2\gamma\cos\theta\sin\phi_0(x)-\mu_1\sin^2\phi_0(x)
\right)=0, 
\end{eqnarray}
which yields 
\begin{eqnarray}
\mu_1=\frac{\delta(1+\delta^2)\coth\pi\delta}{\gamma\cos\theta(1+4\delta^2)}.
\end{eqnarray}
Using the relation $\tan\theta=-\delta+\sigma\mu_1$, we obtain
(\ref{eq:sigma}).


\section*{References}

\begin{thebibliography}{10}

\bibitem{mikhailov06}
A.~S. Mikhailov and K.~Showalter.
\newblock Control of waves, patterns and turbulence in chemical systems.
\newblock {\em Phys. Rep.}, 425:79, 2006.

\bibitem{vanag00}
Vladimir~K. Vanag, Lingfa Yang, Milos Dolnik, Anatol~M. Zhabotinsky, and
  Irving~R. Epstein.
\newblock Oscillatory cluster patterns in a homogeneous chemical system with
  global feedback.
\newblock {\em Nature}, 406:389, 2000.

\bibitem{vanag00_th_exp}
V.~K. Vanag, A.~M. Zhabotinsky, and I.~R. Epstein.
\newblock Pattern formation in the Belousov-Zhabotinsky reaction with
  photochemical global feedback.
\newblock {\em J. Phys. Chem.}, 104:11566, 2000.

\bibitem{pollmann01}
M.~Pollmann, M.~Bertram, and H.~H. Rotermund.
\newblock Influence of time delayed global feedback on pattern formation in
  oscillatory CO oxidation on Pt(110).
\newblock {\em Chem. Phys. Lett.}, 346:123, 2001.

\bibitem{bertram03-ex}
M.~Bertram, C.~Beta, M.~Pollmann, A.~S. Mikhailov, H.~H. Rotermund, and
  G.~Ertl.
\newblock Pattern formation on the edge of chaos: Experiments with CO oxidation
  on a Pt(110) surface under global delayed feedback.
\newblock {\em Phys.\ Rev.\ E}, 67:036208, 2003.

\bibitem{kim01}
M.~Kim, M.~Bertram, M.~Pollman, A.~von Oertzen, A.~S. Mikhailov, H~.H.
  Rotermund, and G.~Ertl.
\newblock Controlling chemical turbulence by global delayed feedback: Pattern
  formation in catalytic CO oxidation on Pt(110).
\newblock {\em Science}, 292:1357, 2001.

\bibitem{beta03}
C.~Beta, M.~Bertram, A.~S. Mikhailov, H.~H. Rotermund, and G.~Etrl.
\newblock Controlling turbulence in a surface chemical reaction by time-delay
  autosynchronization.
\newblock {\em Phys.\ Rev.\ E}, 67:046224, 2003.

\bibitem{popovych05}
O.~V. Popovych, C.~Hauptmann, and P.~A. Tass.
\newblock Effective desynchronization by nonlinear delayed feedback.
\newblock {\em Phys.\ Rev.\ Lett.}, 94:164192, 2005.

\bibitem{hauptmann05}
C.~Hauptmann, O.~V. Popovych, and P.~A. Tass.
\newblock Delayed feedback control of synchronization in locally coupled
  neuronal networks.
\newblock {\em Neurocomputing}, 65:759, 2005.

\bibitem{popovych06}
O.~V. Popovych, C.~Hauptmann, and P.~A. Tass.
\newblock Control of neuronal synchrony by nonlinear delayed feedback.
\newblock {\em Biological Cybernetics}, 95:69, 2006.

\bibitem{tukhlina07}
N.~Tukhlina, M.~Rosenblum, A.~Pikovsky, and J~Kurths.
\newblock Feedback suppression of neural synchrony by vanishing stimulation.
\newblock {\em Phys.\ Rev.\ E}, 75:011918, 2007.

\bibitem{rosenblum04}
M.~G. Rosenblum and A.~S. Pikovsky.
\newblock Controlling synchronization in an ensemble of globally coupled
  oscillators.
\newblock {\em Phys.\ Rev.\ Lett.}, 92:114102, 2004.

\bibitem{rosenblum04_2}
M.~Rosenblum and A.~Pikovsky.
\newblock Delayed feedback control of collective synchrony: An approach to
  suppression of pathological brain rhythms.
\newblock {\em Phys.\ Rev.\ E}, 70:041904, 2004.

\bibitem{yang00}
L.~Yang, M.~Dolnik, A.~M. Zhabotinsky, and I.~R. Epstein.
\newblock Oscillatory clusters in a model of the photosensitive
  Belousov-Zhabotinsky reaction system with global feedback.
\newblock {\em Phys.\ Rev.\ E}, 62:6414, 2000.

\bibitem{rotstein03}
H.~G. Rotstein, N.~Kopell, A.~M. Zhabotinsky, and I.~R. Epstein.
\newblock Canard phenomenon and localization of oscillations in the
  Belousov-Zhabotinsky reaction with global feedback.
\newblock {\em J. Chem. Phys.}, 119:8824, 2003.

\bibitem{bertram03-the}
M.~Bertram and A.~S. Mikhailov.
\newblock Pattern formation on the edge of chaos: Mathematical modeling of co
  oxidation on a Pt(110) surface under global delayed feedback.
\newblock {\em Phys.\ Rev.\ E}, 67:036207, 2003.

\bibitem{beta04}
C.~Beta and A.~S. Mikhailov.
\newblock Controlling spatiotemporal chaos in oscillatory reaction-diffusion
  system by time-delay autosynchronization.
\newblock {\em Physica D}, 199:173, 2004.

\bibitem{parmananda01}
P.~Parmananda and J.~L. Hudson.
\newblock Controlling spatiotemporal chemical chaos using delayed feedback.
\newblock {\em Phys.\ Rev.\ E}, 64:037201, 2001.

\bibitem{kiss07}
Istvan~Z. Kiss, Craig~G. Rusin, Hiroshi Kori, and John~L. Hudson.
\newblock {``Engineering Complex Dynamical Structures: Sequential Patterns and
  Desynchronization''}.
\newblock {\em Science}, 316:1886, 2007.

\bibitem{kori08}
H.~Kori, C.~G. Rusin, I.~Z. Kiss, and J.~L. Hudson.
\newblock Synchronization engineering: Theoretical framework and application to
  dynamical clustering.
\newblock {\em Chaos}, 18:026111, 2008.

\bibitem{kuramoto84}
Y.~Kuramoto.
\newblock {\em Chemical Oscillations, Waves, and Turbulence}.
\newblock Springer, New York, 1984.

\bibitem{okuda93}
K.~Okuda.
\newblock Variety and generality of clustering in globally coupled oscillators.
\newblock {\em Physica\ D}, 63:424, 1993.

\bibitem{ottino07}
William.~L. Kath and Julio.~M. Ottino.
\newblock {``Rhythm Engineering''}.
\newblock {\em Science}, 316:1857, 2007.

\end{thebibliography}
\end{document}